\documentstyle[aps,prb,multicol,eqsecnum,epsf]{revtex}
\begin{document}
\draft
\title{Pseudogap Formation in an Electronic System with $d$-wave 
Attraction at Low-density}

\author{Takashi Hotta, Matthias Mayr, and Elbio Dagotto}

\address{National High Magnetic Field Laboratory, 
Florida State University, Tallahassee, Florida 32306}

\date{\today}

\maketitle

\begin{abstract}

On the basis of an electronic model with separable attractive interaction,
the precursors at high temperature and strong coupling 
of the $d$-wave superconducting state are investigated in the 
one-particle spectral function $A({\bf k},\omega)$ and 
the total density of states $\rho(\omega)$, 
with the use of the self-consistent $t$-matrix approximation.
In the low-density region, it is found that a gap-like structure 
at the Fermi level appears in $A({\bf k},\omega)$ and 
$\rho(\omega)$ above the superconducting transition temperature.
It is shown that the pseudogap energy scale is determined 
by the binding energy of the Cooper-pair.

\end{abstract}


\begin{multicols}{2}
\narrowtext

\section{Introduction}
\label{sec:1}

In the underdoped high-$T_c$ superconductors (HTSC), pseudogap (PG) 
behavior has been widely observed in experiments such as 
NMR, \cite{nmr} specific heat,\cite{specific} and photoemission.\cite{arpes}
All these phenomena can be basically understood as caused by the suppression 
of low-energy spectral weight in the temperature range 
$T_c  \alt T \alt \Delta_{\rm PG}$,
where $T_c$ is the superconducting transition temperature and 
$\Delta_{\rm PG}$ is a characteristic energy scale for the PG formation.
This occurs both in the spin- and
charge-excitation spectra.
As a consequence, the problem is reduced to the clarification  of 
the origin of this spectral-weight suppression,
namely, the physical origin of $\Delta_{\rm PG}$.

The angle-resolved photoemission (ARPES) spectrum, 
which is sensitive to the momentum dependence of the PG,
has revealed that the PG phenomenon itself 
exhibits a $d$-wave symmetry which is smoothly connected to the 
$d$-wave superconducting gap.\cite{arpes}
Moreover, the locus of the minimum gap position in momentum space 
traces the shape of the Fermi surface. From these results, 
it can be inferred that the energy scale for PG formation 
is closely related to the superconducting correlation. 
Then, one of the possible explanations for the PG behavior involves the
discussion of possible ``precursors'' of the Cooper-pair formation 
above $T_c$. 
Certainly there are other possible scenarios that also lead to PG 
formation such as spinon-pairing,\cite{rvb} 
antiferromagnetic spin fluctuation,\cite{af}
and fermion-boson model,\cite{fbm}
but in this paper the focus will be precisely on the development of 
a PG at strong coupling due to the formation of electron bound-states 
at a temperature scale larger than the one corresponding to 
long-range superconducting pairing.

Along this scenario, much effort has been devoted to the investigation
of the PG phenomena.
\cite{pg1,pg2,pg3,pg4,pg5,pg6,pg7,pg8,pg9,pg10,pg11,pg12,pg13,pg14,pg15,pg16}
However, there are few results in the literature leading to PG with $d$-wave 
symmetry, while the PG in the $s$-wave superconductor has been
intensively investigated on the basis of the negative-$U$ Hubbard model.
The popularity of the $s$-wave calculations as opposed to the more
realistic $d$-wave case is mainly due to technical issues.
The quantum Monte Carlo (QMC) simulation provides accurate
information on the negative-$U$ Hubbard model and with these results
the validity of other diagrammatic method such as the self-consistent 
$t$-matrix approximation (SCTMA) can be checked.
However, for the model with $d$-wave attraction (or the nearest-neighbor 
attraction), QMC calculations are difficult mainly due to sign-problems
in the simulations and also because phase 
separation could occur for a model with an 
attractive potential that acts at finite distances, contrary to what
occurs in the attractive Hubbard model where the attraction is only on-site.
\cite{dagotto1}

In spite of these potential difficulties, here the $d$-wave PG is studied 
in order to contribute to the investigation of the energy scale 
$\Delta_{\rm PG}$ in HTSC.
For this purpose, here an effective model with $d$-wave separable 
attraction is analyzed, focussing our efforts into the low-density regime,
for the following reasons.
First, from a physical point of view, 
the low carrier density region is important because 
the underdoped HTSC regime as a first approximation can be described 
as a low-density gas of holes in an antiferromagnetic background. Previous
numerical studies have shown that holes in such an environment behave 
like quasiparticles with a bandwidth renormalized to be of order $J$, 
the Heisenberg exchange
coupling.\cite{nazarenko}
Second, now from a technical viewpoint, it is known that 
the SCTMA gives reliable results in the dilute limit.\cite{tmatrix}
Then, the behavior of the spectral function can be safely investigated 
in the low-density region. For these reasons
in the present paper the average electron filling will actually be at
most $10\%$. Note that our ``electrons'' below will simply represent
fermions interacting through an attractive potential, and thus they can
be thought of as ``holes'' in the context of HTSC.

As mentioned before, the preformed $s$-wave pairing features  
in the negative-$U$ 
Hubbard model have been widely investigated as a prototype for PG formation
in the underdoped HTSC. Besides the technical aspects already discussed,
this seems to be based on the assumption that 
the difference $s$ vs $d$ in the pairing symmetry 
does not play an essential role in the PG formation.
This may seem correct by observing the gap-like structure in the total 
density of states (TDOS), because it appears around the 
Fermi level irrespective of the pairing symmetry, 
although the actual detailed shape is different.
However, recalling that the main features for the PG formation
in the underdoped HTSC have been revealed using ARPES technique,
the structure in the individual one-particle spectral function should
play a crucial role.
In fact, important differences between $s$- and $d$-wave symmetry fairly 
clearly appear in the spectral function described below in our study.

In this paper, it is reported that the $\Delta_{\rm PG}$ scale
agrees with the binding energy of the Cooper-pair irrespective of the 
pairing symmetry.
The main difference between $s$- and $d$-wave
symmetry, appears in the momenta of preformed-pair electrons, 
${\bf K}$ and ${\bf -K}$.
For the $s$-wave symmetry, ${\bf K}$ is always determined by the
band structure.  Namely, in the dilute limit, it is given by the momentum
at the bottom of the band, ${\bf k^*}$.
Since the attraction is uniform in momentum space, 
${\bf K}$ is determined only by the kinetic energy for the $s$-wave case.
On the other hand, for a strong attraction with $d$-wave symmetry,
${\bf K}$ is not given by ${\bf k^*}$, but is located at 
$(\pi,0)$ and $(0,\pi)$, because the attraction becomes maximum
at those momenta. 
Such a competition between the band structure and the strong attractive
interaction leads to interesting features in the 
$d$-wave PG, while the $s$-wave PG simply follows the band structure.

This paper is organized as follows. In section II, a general
formalism to calculate the electronic self-energy in the SCTMA
on the real-frequency axis is present.
For the investigation of the PG structure for the $d$-wave attraction, 
a technical trick called ``the $s$-$d$ conversion'' is introduced. 
Section III is devoted to the results obtained with the formalism of 
Sec.~II.
Two types of band structures are considered with ${\bf k^*}=(0,0)$
and $(\pi,0)$, respectively.
In section IV, the results are discussed.
Finally in section V, after providing some comments, the main results of
this paper are summarized.
Throughout this paper, units such that $\hbar=k_{\rm B}=1$ are used.

\section{Formulation}
\label{sec:2}

\subsection{Hamiltonian}

Let us consider a simple model in which electrons are coupled with
each other through a separable attractive interaction.
The symmetry of the electron pair is contained in the 
attractive term of the model, but it is not necessary to write it 
explicitly in most of the formulation of this section, although  
it will become important for the discussion on the PG.
The model Hamiltonian is written as
\begin{eqnarray}
  \label{hamiltonian}
  H &=& \sum_{{\bf k}\sigma} (\varepsilon_{\bf k}-\mu)
  c_{{\bf k}\sigma}^{\dag}c_{{\bf k}\sigma}  \nonumber \\
  &+& \sum_{\bf k,k',q}V_{\bf k,k'}
  c_{{\bf k}\uparrow}^{\dag}c_{{\bf k+q}\uparrow}
  c_{{\bf k'}\downarrow}^{\dag}c_{{\bf k'-q}\downarrow},
\end{eqnarray}
where $c_{{\bf k}\sigma}$ is the annihilation operator for
an electron with momentum ${\bf k}$ and spin $\sigma$,
$\varepsilon_{\bf k}$ the one-electron energy,
$\mu$ the chemical potential, and 
$V_{\bf k,k'}$ the pair-interaction between electrons.
The electron dispersion is expressed as 
\begin{equation}
  \varepsilon_{\bf k}= -2t (\cos k_x + \cos k_y)-4t'\cos k_x \cos k_y,
\end{equation}
where $t$ and $t'$ are the nearest and next-nearest
neighbor hopping amplitudes, respectively.
The pair-interaction is written as
\begin{equation}
  V_{\bf k,k'}= -V f_{\bf k}f_{\bf k'},
\end{equation}
where $f_{\bf k}$ is the form factor characterizing the symmetry
of the Cooper-pair. 
Note that a positive value of $V$ denotes an attractive 
interaction throughout this paper.

\subsection{Self-consistent $t$-matrix approximation}

Now let us calculate the spectral function using the SCTMA.
Since this method becomes exact in the two-particle problem,
it is expected to give a reliable result in the low-density region.
\cite{tmatrix}
In fact, this expectation has been already checked in the attractive Hubbard 
model by comparing SCTMA results against QMC simulations.\cite{randeria}
Therefore the reliability of the SCTMA may also be expected for the 
non-$s$-wave attractive interaction, 
even though the direct comparison with QMC results is quite difficult
in this case.

\begin{figure}[h]
\hskip-0.6truein
\centerline{\epsfxsize=2.0truein \epsfbox{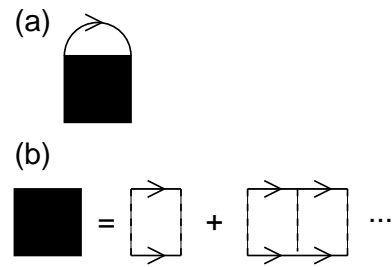} } 
\vskip-1.5truein
\caption{(a) Self-energy in the SCTMA. The hatched square and 
the solid line denote the $t$-matrix $\Gamma$ and the renormalized
Green's function $G$, respectively.
(b) Diagrammatic representation for $\Gamma$.
The broken lines indicate the interaction.}
\label{fig1}
\end{figure}

Consider first for completeness the imaginary-axis representation.
In this formulation, the one-particle Green's function $G$ is
given by
\begin{equation}
 \label{gfnim}
  G({\bf k},i\omega_n)
  = {1 \over i\omega_n-(\varepsilon_{\bf k}-\mu)-
    \Sigma({\bf k},i\omega_n)},
\end{equation}
where $\omega_n=\pi T (2n+1)$, $n$ is an integer, 
and $T$ the temperature.
In the SCTMA, the self-energy $\Sigma({\bf k},i\omega_n)$ is
obtained with the use of the $t$-matrix given by the infinite sum of 
particle-particle (p-p) ladder diagrams, as shown in Fig.~1.
More explicitly, $\Sigma$ is expressed as 

\begin{eqnarray}
 \label{selfim}
  \Sigma({\bf k},i\omega_n)
  &=& f_{\bf k}^2 ~T \sum_{n'}\sum_{\bf k'} 
  \Gamma({\bf k+k'},i\omega_n+i\omega_{n'}) \nonumber \\
  &\times& G({\bf k'},i\omega_{n'}),
\end{eqnarray}
where $\Gamma({\bf q}, i\nu_m)$ is the $t$-matrix, given by
\begin{equation}
  \label{tmatrixim}
  \Gamma({\bf q},i\nu_m)
  = {-V^2\phi({\bf q},i\nu_m) \over 1-V \phi({\bf q},i\nu_m)}.
\end{equation}
Here $\nu_m=2\pi T m$, with $m$ an integer, and 
$\phi({\bf q},i\nu_m)$ is the p-p ladder, defined by 
\begin{eqnarray}
\label{ppim}
  \phi({\bf q},i\nu_m)
  &=& T \sum_ n \sum_{\bf k} f_{\bf k}^2 G({\bf k},i\omega_n) \nonumber \\
  &\times&   G({\bf q-k},i\nu_m-i\omega_n).
\end{eqnarray}
Note that the Hartree term is neglected in the self-energy 
because it should be considered as included in the band structure.
The Green's function can be calculated self-consistently
using Eqs.~(\ref{gfnim})-(\ref{ppim}).
The chemical potential is determined by
\begin{equation}
  \langle n \rangle /2= T \sum_ n \sum_{\bf k} e^{i\omega_n \eta}
  G({\bf k},i\omega_n),
\end{equation}
where $\langle n \rangle$ is the average electron number density per site,
and $\eta$ is an infinitesimal positive number. 
In order to obtain results on the real-frequency axis, 
Pad\'e approximants for the numerical
analytic continuation from the imaginary-axis data
are frequently used.\cite{pade}
However, in general, it is difficult to control the accuracy of 
the calculation by this procedure. 

In this paper, our efforts are focused on the
direct calculation of the Green's function on the real-frequency 
axis.\cite{pg12}
In this context, a self-consistent calculation
for the spectral function
\begin{equation}
A({\bf k},\omega) =(-1/\pi){\rm Im}G({\bf k},\omega),
\end{equation}
is carried out, where the retarded Green's function $G({\bf k},\omega)$ 
is given by
\begin{equation}
  G({\bf k},\omega)
  ={1 \over \omega-(\varepsilon_{\bf k}-\mu)-\Sigma({\bf k},\omega)}.
\end{equation}
The imaginary part of the retarded self-energy is expressed as
\begin{eqnarray}
  \label{self}
  {\rm Im}\Sigma ({\bf k},\omega) &=&
   f_{\bf k}^2 \sum_{\bf k'} \int d\omega'
  [f_{\rm F}(\omega')
  +f_{\rm B}(\omega+\omega')] \nonumber \\
  &\times & A({\bf k'}, \omega') {\rm Im}\Gamma({\bf k+k'},\omega+\omega'),
\end{eqnarray}
where $f_{\rm F}(x)=1/(e^{x/T}+1)$ and $f_{\rm B}(x)=1/(e^{x/T}-1)$.
The real-part of $\Sigma$ is obtained through the use of ${\rm Im}\Sigma$
in the Kramers-Kronig (KK) relation 
\begin{equation}
  {\rm Re}~\Sigma({\bf k},\omega)
  = {\rm p.v.}\int {d\omega' \over \pi}
  {{\rm Im}~\Sigma({\bf k},\omega') \over \omega-\omega'},
\end{equation}
where p.v. means the principal-value integral. 
The $t$-matrix is
\begin{equation}
 \Gamma({\bf q},\omega)
  = {-V^2 \phi({\bf q},\omega) \over 1-V \phi({\bf q},\omega)},
\end{equation}
where ${\rm Im}\phi({\bf q},\omega)$ is given by
\begin{eqnarray}
  {\rm Im}~\phi({\bf q},\omega)
  &=& \pi \sum_{\bf k}\int d\omega' f_{\bf k}^2 
  ~{\rm tanh}{\omega' \over 2T}
  A({\bf k}, \omega')  \nonumber \\
 &\times& A({\bf q-k},\omega-\omega'),
\end{eqnarray}
and the real part of $\phi({\bf q},\omega)$ is also obtained 
using the KK relation.
The electron number is obtained through
\begin{equation}
  \langle n \rangle /2= \sum_{\bf k} \int d\omega
	A({\bf k}, \omega)f_{\rm F}(\omega),
\end{equation}
and the spectral function must satisfy the sum-rule
\begin{equation}
  1= \sum_{\bf k} \int d\omega A({\bf k}, \omega).
\end{equation}
This will be a check for the accuracy of the numerical results presented here.

In the actual calculation, the fast Fourier transformation 
is applied to accelerate the procedure.\cite{fft}
The first Brillouin zone is divided into a $64 \times 64$ lattice
and the frequency integration is replaced by a discrete sum
in the range $-25t < \omega < 25t$, dividing it into $512$ small
intervals.
As a consequence, the energy resolution is about $0.1t$, 
indicating the order of magnitude of the lowest temperature 
at which our calculations can be reliably carried out.
When the relative difference between two successive iterations for 
$A({\bf k},\omega)$ is less than $0.01$ at each $({\bf k},\omega)$, 
the iteration loop is terminated. 
As for the sum-rule, it is systematically found to be satisfied 
within $1\%$.
This value is the limitation for the accuracy of the present 
calculation, because the integral equation with a singular 
kernel is solved 
by replacing the integration procedure 
by a simple discrete summation.

\subsection{$s$-$d$ conversion}

For a separable potential with $d$-wave symmetry, $f_{\bf k}$
is given by
\begin{equation}
f_{\bf k}= \cos k_x -\cos k_y.
\end{equation}
In this case, due to the prefactor $f_{\bf k}^2$ in Eq.~(\ref{self}),
$\Sigma$ always vanishes along the lines $k_x = \pm k_y$,
leading to a delta-function contribution in the spectral function.
In order to avoid this singularity, a self-consistent calculation 
for the $d$-wave case was first attempted by imposing anti-periodic and 
periodic boundary conditions for the $k_x$- and $k_y$-directions, 
respectively. 
However, it was not always possible to obtain a converged self-consistent 
solution in this case.
Actually, it was quite difficult to control such convergence even if the 
temperature $T$ was slowly decreased from the high-temperature region 
in which a stable solution was obtained, 
or if the coupling $V$ was adiabatically 
increased from the weak-coupling region. 

This difficulty is caused by the fact that $\Sigma$ becomes negligibly 
small in the region around $k_x \approx \pm k_y$ for the $d$-wave case,
even if the strong-coupling value for $V$ is set as high as $V=8t$. 
If ${\rm Im}\Sigma$ becomes smaller that the energy resolution in the 
present calculation, which is about $0.1t$, the sharp peak structure in the 
spectral function around $k_x \approx \pm k_y$ is not correctly included 
in the self-consistent calculation.
This leads to a spurious violation of the sum-rule, indicating that technical
problems appear in reaching a physically meaningful solution for $d$-wave 
symmetry at low temperatures.

In order to avoid this difficulty, a continuous change from $s$- to $d$-wave 
symmetry is here considered by introducing a mixing parameter $\alpha$ 
such that 
\begin{equation}
 f_{\bf k}^2 = (1-\alpha) + \alpha (\cos k_x -\cos k_y)^2.
\end{equation}
Our calculations start at $\alpha=0$, i.e., for the pure $s$-wave case,
in which a stable solution can be obtained easily in the SCTMA.
Then, $\alpha$ is gradually increased such that the $d$-wave case is 
approached.
If a physical quantity for the $d$-wave model is needed, 
an extrapolation is made by using the calculated results for the
quantities of interest between $0 \leq \alpha < 1$.

\section{Results}
\label{sec:3}

In this section, our results calculated with the use of the real-axis
formalism are shown. 
Here the magnitude of the interaction $V$ is fixed as $V=8t$.

\subsection{Case of $t'=0$}

Let us consider first the band structure with ${\bf k^*}=(0,0)$.
In Fig.~2(a), the total density of states $\rho(\omega)$ is shown, given by
$\rho(\omega)=\sum_{\bf k} A({\bf k},\omega)$. 
The whole curve for TDOS is not presented
in this figure, because its shape at a larger scale
is quite similar to the non-interacting case.
At $\alpha=0$, a gap-like feature at the Fermi level can be observed,
although it is shallow.
This result has been already reported in numerous previous papers using
several techniques.\cite{randeria}
With the increase of $\alpha$, the gap structure gradually 
becomes narrower and at the same time deeper.
The TDOS extrapolated to $\alpha=1$ using the results for the $\alpha$s in
the figure  is not shown, 
because it becomes unphysically negative in some energy region.
However, this is not a serious problem, because such a behavior is only 
an artifact due to the extrapolation using a small number of $\alpha$-results
and it will disappear if $\alpha$ approaches
unity very slowly and calculations with higher-energy resolution 
are performed. 
This problem is not present in the studies at $t'=0$ in the next subsection.
Thus, this is a small complication that can be solved with
more CPU and memory-intensive studies than reported here.

\bigskip
\begin{figure}[h]
\centerline{\epsfxsize=3.0truein \epsfbox{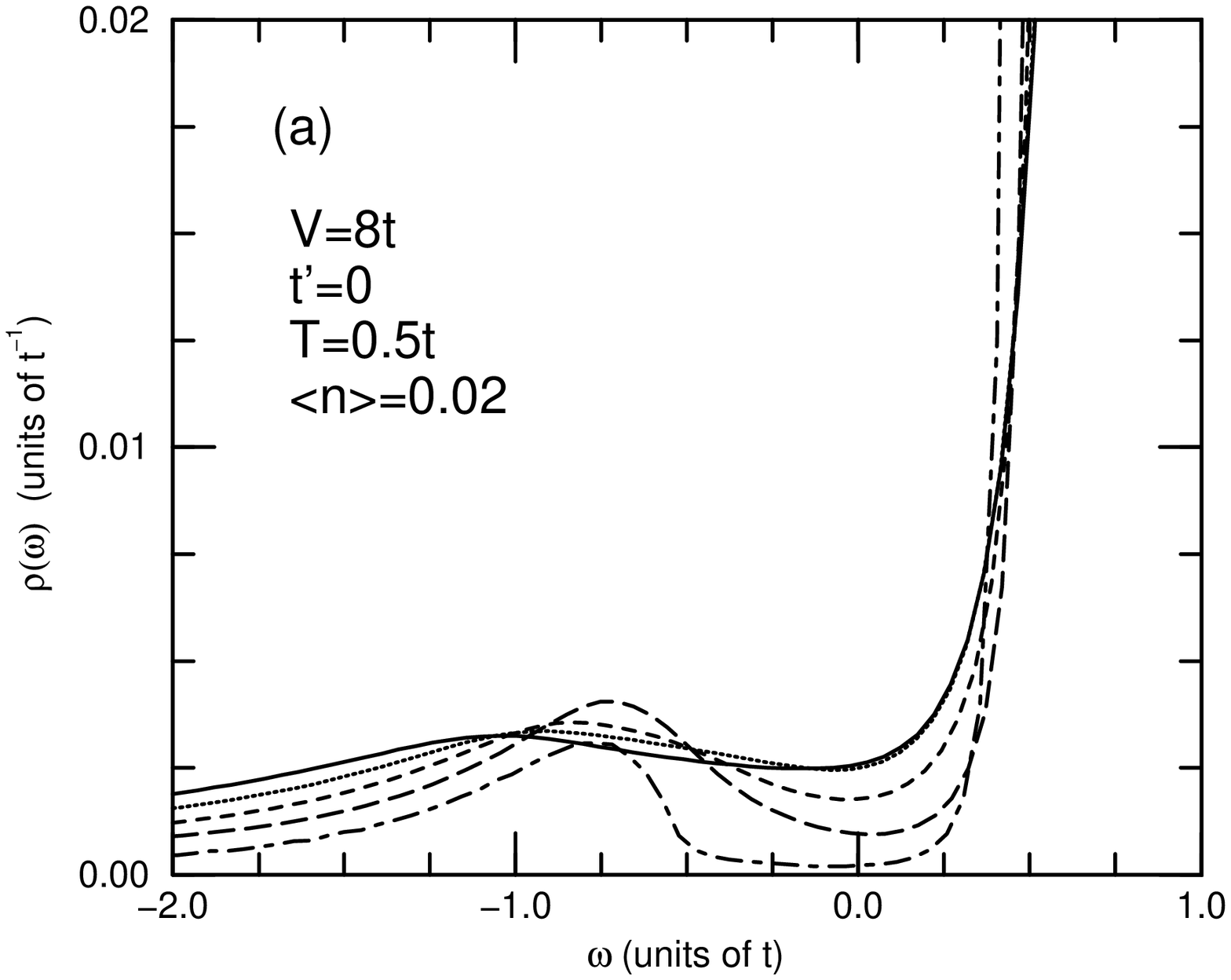} }

\centerline{\epsfxsize=3.0truein \epsfbox{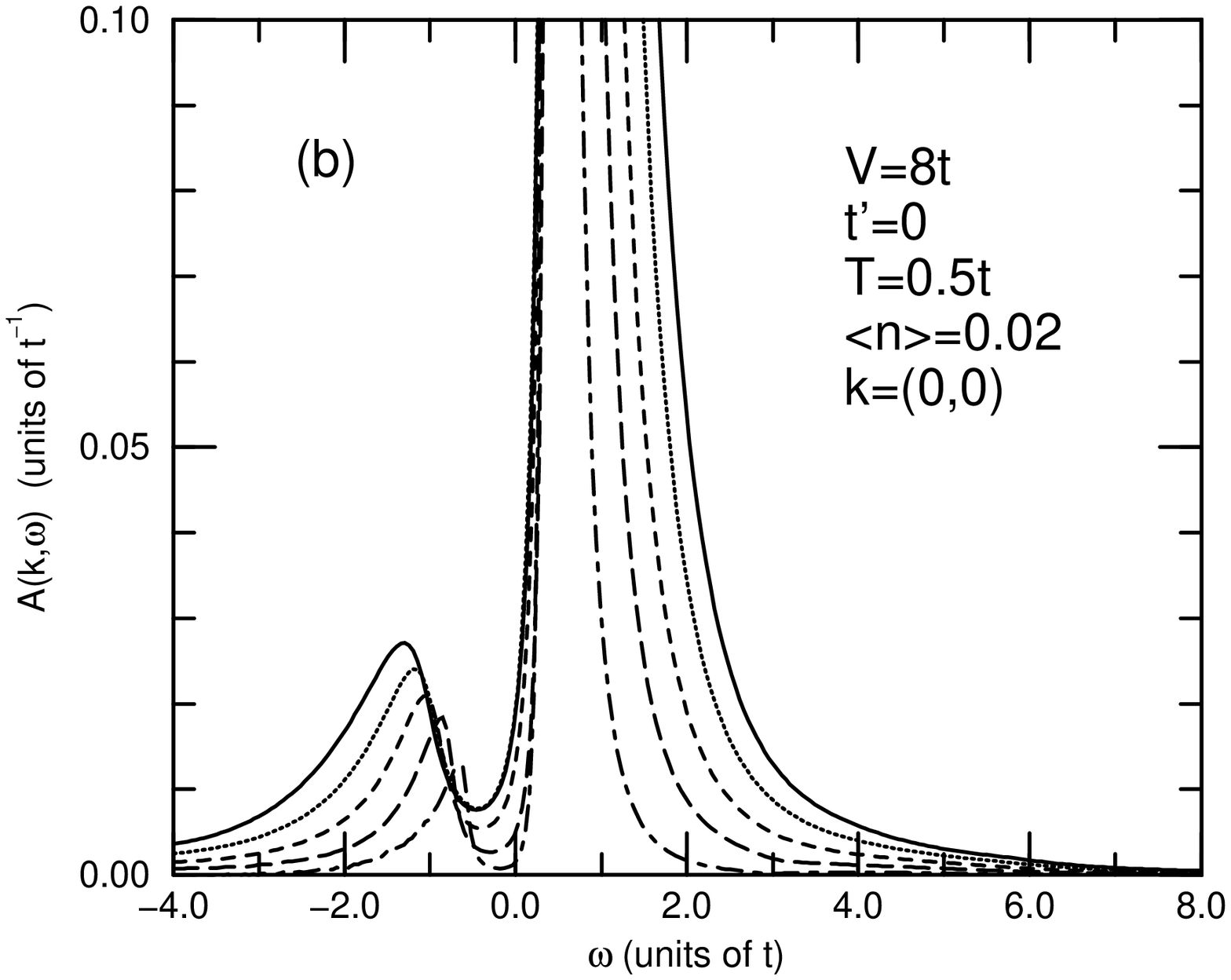} }

\centerline{\epsfxsize=3.0truein \epsfbox{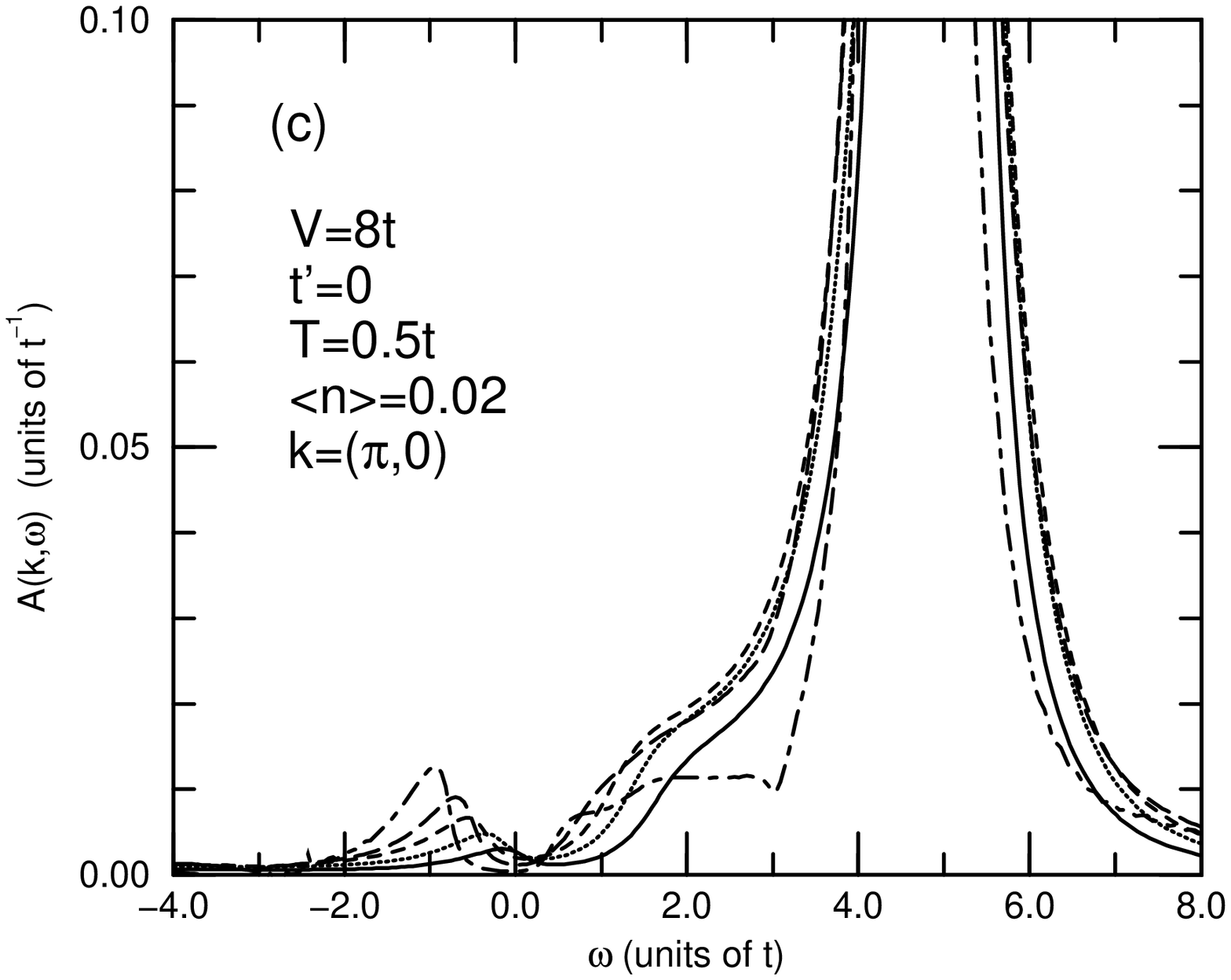} } 

\vskip-0.4truein
\caption{
(a) Total density of states (TDOS) for several values of $\alpha$, the
interpolating parameter between $s-$ and $d$-wave models,
in the case $V=8t$, $t'=0$, $T=0.5t$, and $\langle n \rangle = 0.02$.
The solid, dotted, dashed, long-dashed, and dot-dashed curves
denote the TDOS results for $\alpha=0,0.2,0.4,0.6$, and $0.8$, respectively.
(b) The spectral function at ${\bf k}=(0,0)$.
(c) The spectral function at ${\bf k}=(\pi,0)$.
}
\end{figure}

In order to understand the observed changes in the PG behavior 
with the increase of $\alpha$, special attention must be given to 
the spectral function $A({\bf k},\omega)$.
Let us first analyze the result at ${\bf k}=(0,0)$, 
shown in Fig.~2(b),
in which two peaks are observed.
The large peak above the Fermi level is due to the quasi-particle (QP)
contribution, because if the interaction is 
gradually decreased, it continuously changes into the expected
non-interacting $\delta$-function peak. 
Thus here it will be called ``the QP peak''. However, note that
another structure can be observed below the Fermi level, 
although it has only a small weight. 
As will be discussed in the next subsection, this originates from 
the peak structure in ${\rm Im}\Gamma$.
In this sense, it can be called ``the resonant peak'' due to the 
formation of the bound pair.\cite{pg4}
When $\alpha$ is increased, the QP peak becomes sharper and 
the position of the resonant peak is shifted to the right side in Fig.~2(b),
while the weight decreases.
At $\alpha=1$, the resonant peak will likely disappear and only 
the $\delta$-function QP peak will occur, since the self-energy vanishes 
due to the prefactor $f^2_{\bf k}$ in Eq.~(\ref{self}).

Although the weight for the resonant peak in $A({\bf k},\omega)$ with
${\bf k}=(0,0)$ decreases with the increase of $\alpha$, it is actually 
transfered to another $A({\bf k},\omega)$ with ${\bf k} \ne (0,0)$.
Then,  let us next turn our attention to $A({\bf k},\omega)$ with 
${\bf k}=(\pi,0)$, shown in Fig.~2(c).
In this case, a QP peak is also observed, 
but the position is higher than that at ${\bf k}=(0,0)$.
The difference between the positions of those QP peaks is about $4t$,
namely, equal to $\varepsilon_{(\pi,0)}-\varepsilon_{(0,0)}$.
It should be noted that another peak structure grows below 
the Fermi level with the increase of $\alpha$.
The position roughly agrees with the lower edge of 
the PG structure in the TDOS, suggesting that the PG structure 
for $d$-wave originates from 
$A({\bf k},\omega)$ around ${\bf k}=(\pi,0)$.

Let us summarize this subsection. 
Pseudogap features appear in the density of states both for $s$- and
$d$-wave models, 
but its origin is quite different.
For the $s$-wave case, this structure is mainly due to the preformed 
pair of electrons around the point ${\bf k}={\bf k^*}=(0,0)$ at the
bottom of the band.
On the other hand, for the $d$-wave case, it originates from the 
pair of electrons at other ${\bf k}$-points, 
especially, ${\bf k}=(\pi,0)$.
In the case of strong attraction such as $V=8t$, those electrons
can exploit the effect of the attractive interaction,
in spite of the loss of the kinetic energy. 
In other word, this difference is due to the competition between the 
kinetic and the interaction effects.

\subsection{Case of $t' \ne 0$}

 From the result for $t'=0$, in order to obtain a large PG structure 
for $d$-wave symmetry, it is necessary to consider the band structure in 
which ${\bf k^*}$'s are located at $(\pm \pi,0)$ and $(0,\pm \pi)$.
The reason is that electrons around ${\bf k}={\bf k^*}$ can exploit 
the kinetic as well as the pairing energy due to the strong 
attractive interaction.
As for a value of $t'$, it is here typically chosen as $t'=-t$ but the
results do not depend crucially on such a choice.

\vskip1cm

\begin{figure}[h]
\centerline{\epsfxsize=3.0truein \epsfbox{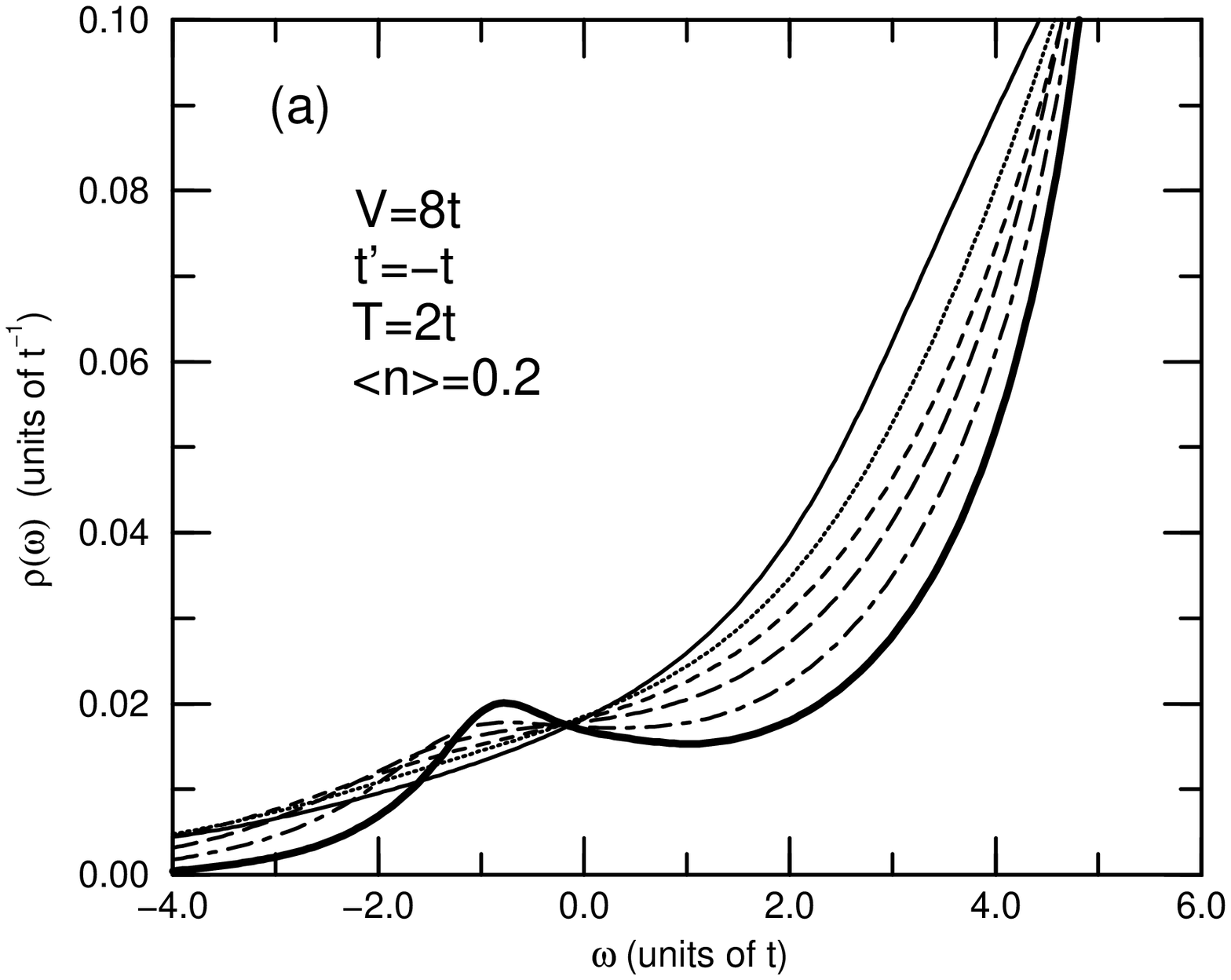} }

\centerline{\epsfxsize=3.0truein \epsfbox{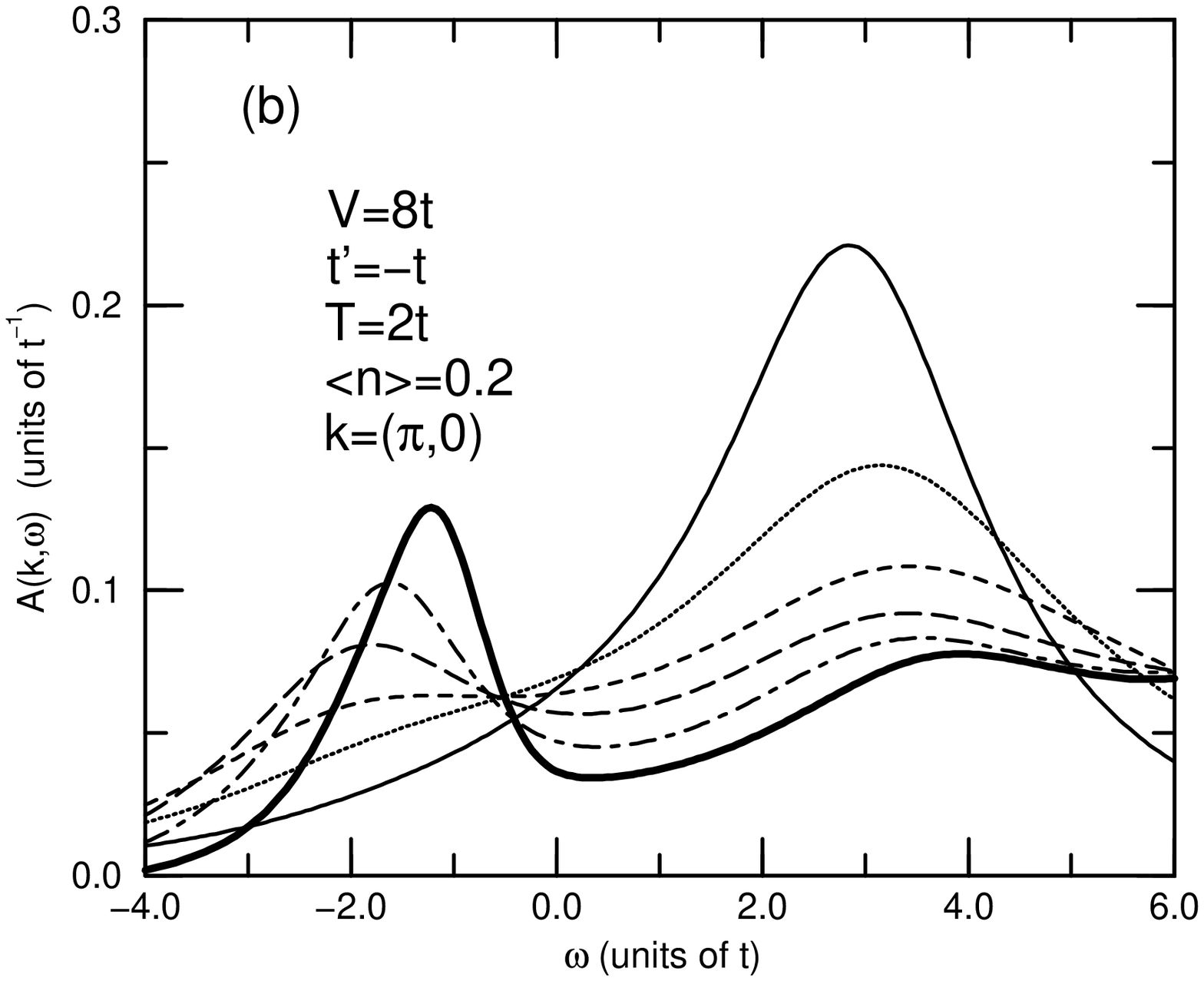} }

\centerline{\epsfxsize=3.0truein \epsfbox{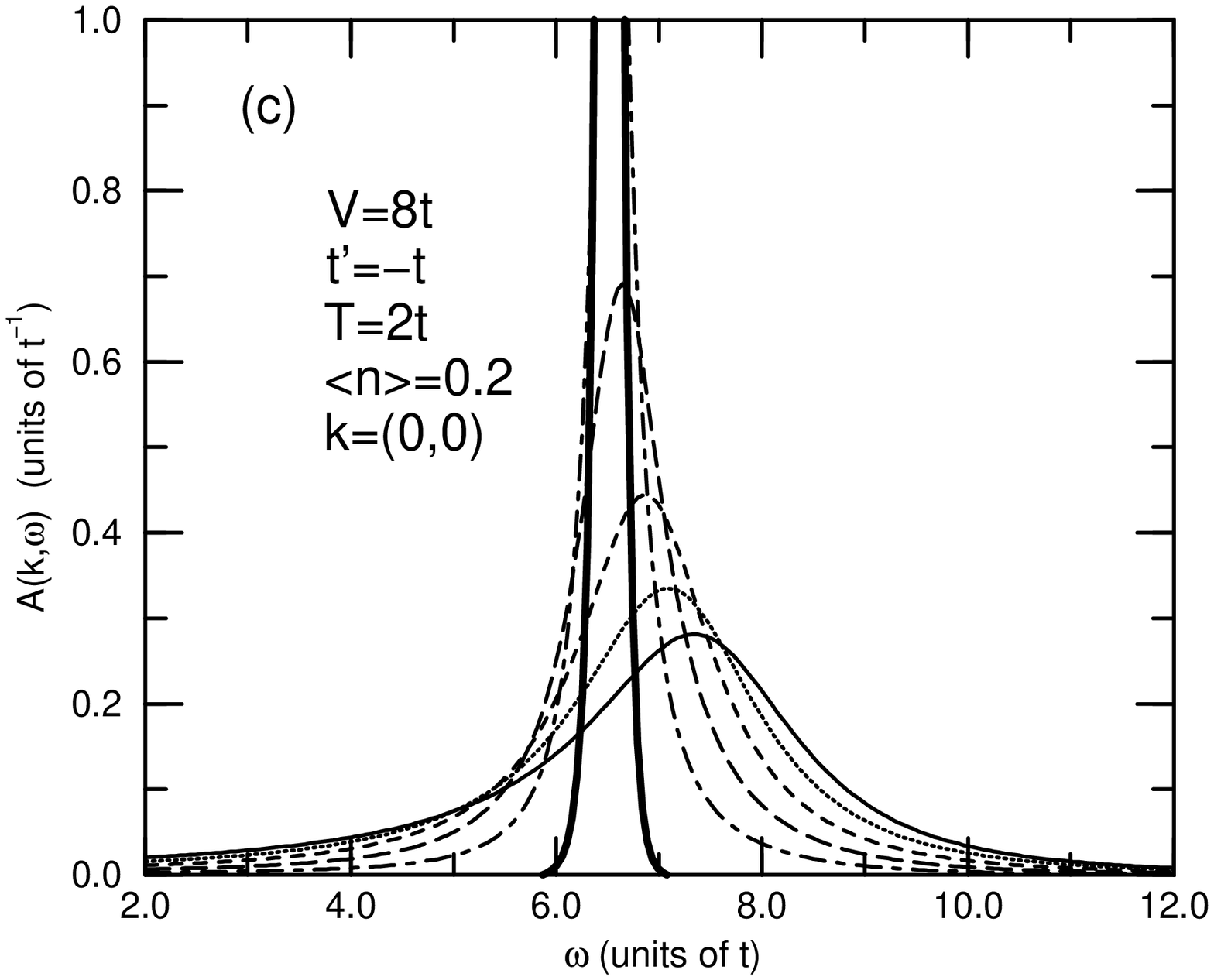} } 

\vskip-0.4truein

\caption{
(a) Total density of states (TDOS) for several values of $\alpha$
in the case of $V=8t$, $t'=-t$, $T=2.0t$, and $\langle n \rangle=0.2$.
The solid, dotted, dashed, long-dashed, and dot-dashed
curves denote the TDOS for $\alpha=0,0.2,0.4,0.6$, and $0.8$, respectively.
The thick solid curve is an extrapolated to $\alpha=1$ result from the
available  TDOS's
in the interval $0 \leq \alpha <1$.
(b) The spectral function at ${\bf k}=(\pi,0)$.
(c) The spectral function at ${\bf k}=(0,0)$.
}
\label{fig3}
\end{figure}

\par\vfill
\eject

In the TDOS shown in Fig.~3(a), no structure around the Fermi level
is observed for the $s$-wave case, but a peak appears below the 
Fermi level with the increase of $\alpha$.
It can be regarded as a sign of PG formation, 
but this interpretation becomes much clearer if $A({\bf k},\omega)$
is investigated.
In Fig.~3(b), the change of $A({\bf k},\omega)$ at 
${\bf k}={\bf k^*}=(\pi,0)$ is depicted when $\alpha$ is increased. 
In the pure $s$-wave case, a large QP peak can be observed, but
it is difficult to find a resonant peak below the Fermi level.
On the other hand, when the $d$-wave is approached 
by increasing $\alpha$,
the QP peak is gradually destroyed and the resonant peak grows strongly 
below the Fermi level.
Then, the PG structure is much larger compared to that at $t'=0$.
Note that in this case, the extrapolation for the TDOS at $\alpha=1$ 
is quite successful, contrary to what occurs at $t'=0$,
 because the large size of the PG allows 
us to perform 
the calculation at a high temperature such as $T=2t$, a situation in which
the structure in the TDOS is 
smoother than at $t'=0$.

For the case $t'=-t$, weight transfer in 
$A({\bf k},\omega)$ is observed with the increase of $\alpha$,
but it occurs between the QP and the resonant peaks at ${\bf k}=(\pi,0)$.
In order to confirm this idea, $A({\bf k},\omega)$ at 
${\bf k}=(0,0)$ was studied as shown in Fig.~3(c).
As expected, only the sharpening of the QP peak is observed as $\alpha$
is varied, 
because the strength of the attractive interaction at 
${\bf k}=(0,0)$ becomes weak with the increase of $\alpha$.
Note that a finite width for the QP remains at $\alpha=1$, but it is 
only a numerical artifact.
Actually at  $t'=-t$, electrons around ${\bf k}=(0,0)$ do not 
take part in the PG formation even for the $s$-wave case.
Since electrons around ${\bf k}={\bf k^*}$ gain both the kinetic 
and potential energies, the PG structure is determined only by those
electrons.

\section{Discussion}

\subsection{Energy scale for pseudogap}

 From the results in the previous section, in addition to the QP peak, 
a resonant peak below the Fermi level in $A({\bf k^*},\omega)$ 
has been observed, although its appearance depends on the value of 
$t'$ and the symmetry of the pair interaction.
These two peaks define the PG structure in $A({\bf k^*},\omega)$ and 
also in the TDOS, although in the latter it is often difficult to 
observe due to the smearing effects of the sum over momentum of the
individual one-particle spectral functions.
Based on these observations, 
in this paper the PG energy $\Delta_{\rm PG}$ is defined by the 
width between the QP and the resonant peaks in $A({\bf k^*},\omega)$.
Note that for $t'=0$ and $\alpha=1$, the weight for the resonant peak 
in $A({\bf k^*},\omega)$ will vanish, but in the limit of 
$\alpha \rightarrow 1$, its position approaches the lower peak of 
the PG structure.

In order to elucidate the physical meaning of our $\Delta_{\rm PG}$,
the imaginary part of the self-energy is analyzed
at ${\bf k}={\bf k^*}$, because its structure has a direct effect on
the spectral function, given by
\begin{eqnarray}
 && A({\bf k},\omega) \nonumber \\ 
 && =-{1 \over \pi}{{\rm Im}\Sigma({\bf k},\omega) \over 
   [\omega-(\varepsilon_{\bf k}-\mu)-{\rm Re}\Sigma({\bf k},\omega)]^2
   +[{\rm Im}\Sigma({\bf k},\omega)]^2}. 
\end{eqnarray}
For an intuitive explanation, it is not convenient to analyze
the full self-consistent solution for ${\rm Im}\Sigma({\bf k},\omega)$.
Rather the essential information can be obtained by simply evaluating 
Eq.~(\ref{self}) replacing the renormalized Green's function $G$ with 
the non-interacting Green's function $G_0$.
Then, $A({\bf k'},\omega)$ on the right-hand side of Eq.~(\ref{self})
becomes $\delta(\omega-\varepsilon_{\bf k'}+\mu)$
and ${\rm Im}\Gamma$ is obtained with the use of the p-p ladder diagrams 
composed of two $G_0$-lines.
Furthermore, only the contribution from the preformed pair with 
momentum zero for the center of mass is considered.
Namely, only ${\rm Im}\Gamma$ with ${\bf k+k'=0}$ is taken into account
in Eq.~(\ref{self}). 

Due to the above simplifications, ${\rm Im}\Sigma$ at ${\bf k=k^*}$
can be shown to be
\begin{eqnarray}
{\rm Im} \Sigma ({\bf k^*},\omega) &\approx& 
f_{\bf k^*}^2 [f_{\rm F}(\varepsilon_{\bf k^*}-\mu)+
f_{\rm B}(\omega+\varepsilon_{\bf k^*}-\mu)] \nonumber \\
&\times& {\rm Im}\Gamma({\bf 0},\omega+\varepsilon_{\bf k^*}-\mu).
\end{eqnarray}
If it is assumed that ${\rm Im}\Gamma$ has a peak at $\omega=\Omega$,
then ${\rm Im} \Sigma ({\bf k^*},\omega)$ shows a peak structure 
around $\omega \approx \Omega-(\varepsilon_{\bf k^*}-\mu)$.
Here the weight of the peak will not be discussed, but
it will have a small finite value if the thermal 
factor is taken into account.
Therefore in the spectral function at ${\bf k}={\bf k^*}$, 
besides the sharp QP peak at $\omega = \varepsilon_{\bf k^*}-\mu$,
another peak appears around 
$\omega \approx \Omega-(\varepsilon_{\bf k^*}-\mu)$ 
due to the peak-structure in 
${\rm Im} \Sigma ({\bf k^*},\omega)$,
indicating that the size of the PG feature is given by
$\Delta_{\rm PG}=|2(\varepsilon_{\bf k^*}-\mu)-\Omega|$
in this simple approximation.

Now let us estimate the value of $\Omega$.
Since $\Omega$ is the energy at which ${\rm \Gamma}T$ 
acquires its maximum value, it can be obtained from the condition
\begin{eqnarray}
 \label{cond}
 1-V {\rm Re}~\phi_0({\bf 0},\Omega)=0,
\end{eqnarray}
where $\phi_0$ is the p-p ladder set with two $G_0$-lines,
explicitly given by
\begin{eqnarray}
 \phi_0({\bf q},\omega)=
 \sum_{\bf k} f^2_{\bf k}
 {f_{\rm F}(\varepsilon_{\bf k}-\mu)
 -f_{\rm F}(-\varepsilon_{\bf q-k}+\mu) \over 
  \omega+i\eta-(\varepsilon_{\bf q-k}+\varepsilon_{\bf k}-2\mu)}.
\end{eqnarray}
In the dilute case in which the chemical potential $\mu$ is situated 
below the lower band-edge $\varepsilon_{\bf k^*}$ 
and in the temperature region for $T \alt \varepsilon_{\bf k^*}-\mu$,
Eq.~(\ref{cond}) reduces to 
\begin{eqnarray}
\label{binding}
1 + V \sum_{\bf k} f_{\bf k}^2 {1 \over \Omega-2(\varepsilon_{\bf k}-\mu)}
=0,
\end{eqnarray}
which is just the equation to obtain the binding energy $\Delta$ 
of the Cooper-pair in the two-particle problem.\cite{Schrieffer}
Since $\Delta$ is defined as the difference between 
the two-particle bound-state energy $\Omega$
and twice the one-particle energy $\varepsilon_{\bf k^*}-\mu$, 
it is given by 
\begin{eqnarray}
\label{delta}
\Delta= 2(\varepsilon_{\bf k^*}-\mu)-\Omega. 
\end{eqnarray}
Then, from this analysis
it is found that $\Delta_{\rm PG}=\Delta$, as intuitively expected.

\subsection{Quantitative comparison between $\Delta$ and $\Delta_{\rm PG}$}

Although the discussion in the previous subsection is too simple 
to address the fully renormalized self-consistent solution, 
the results reported in Sec.~III will become more
 reasonable if the relevant energy scales are
correctly addressed.
In order to understand this, let us make a direct comparison between
the analytic value for $\Delta$ and $\Delta_{\rm PG}$ evaluated from 
the energy difference between the two peaks in $A({\bf k^*},\omega)$.

By solving Eqs.~(\ref{binding}) and (\ref{delta}), the binding energy 
for the $s$- and $d$-wave cases with $t'=0$ and $t'=-t$ is obtained, 
which is shown in Fig.~4(a).
In the strong-coupling region $V \agt 8t$, all curves are proportional 
to $V$. In the weak-coupling region, it is difficult to obtain an 
accurate value numerically, because the binding is exponentially small in 
this region. Especially, for the $d$-wave case with $t'=0$, 
it was not possible to obtain any finite value in a region of 
$V \alt 7t$.
However, when negative $t'$ is introduced, the binding energy for 
$d$-wave pair is much enhanced, while the $s$-wave binding energy is 
not much affected by $t'$.

This result can be understood once again as caused by the competition between 
the band structure and the attractive interaction 
at ${\bf k}={\bf k^*}$. 
For the $s$-wave case, 
since the attractive interaction is isotropic in momentum space,
the $V$ dependence of $\Delta$ is not so sensitive to the 
position of ${\bf k^*}$. 
However, for the $d$-wave symmetry, the situation is drastically different.
For the band structure with ${\bf k^*}=(0,0)$, it is quite difficult
for electrons around ${\bf k}={\bf k^*}$ to form a pair, 
because the attraction does not work at ${\bf k}={\bf k^*}$.
Thus, in the weak-coupling region the binding energy is vanishingly small.
If $V$ becomes as large as the bandwidth, $8t$, 
electron pairs at ${\bf k} \ne {\bf k^*}$ begin to affect
the binding energy and the value of $\Delta$ becomes comparable to $t$.
On the other hand, for the band structure with ${\bf k^*}=(\pi,0)$,
electrons around ${\bf k}={\bf k^*}$ easily form a pair because of 
the large strength of the attraction at that point.
This sensitivity of the $d$-wave binding energy to the band structure is 
consistent with that of the $d$-wave PG observed in the spectral 
function.

Now let us compare our PG energy $\Delta_{\rm PG}$ with $\Delta$. 
In Fig.~4(b), those quantities are depicted as a function of $\alpha$.
Note that these $\Delta_{\rm PG}$'s are estimated from 
$A({\bf k^*},\omega)$'s in Figs.~2(b) and 3(b).
In the region $\alpha <0.4$ for $t'=-t$, the values of
$\Delta_{\rm PG}$ are not shown, 
because the resonant peak could not be observed for
the parameters used in Fig.~3(b).
For the case of $t'=0$, $\Delta_{\rm PG}$ traces the curve of the 
binding energy, though there is a small deviation between them.
On the other hand, for the case of $t'=-t$, the deviation is larger
particularly around $\alpha \approx 0.6$, but $\Delta_{\rm PG}$ approaches 
$\Delta$ at the $d$-wave case. 
Thus, from our analysis, it is clear that the energy scale for 
the PG structure is simply the pair binding energy.

\bigskip
\begin{figure}[h]
\centerline{\epsfxsize=3.0truein \epsfbox{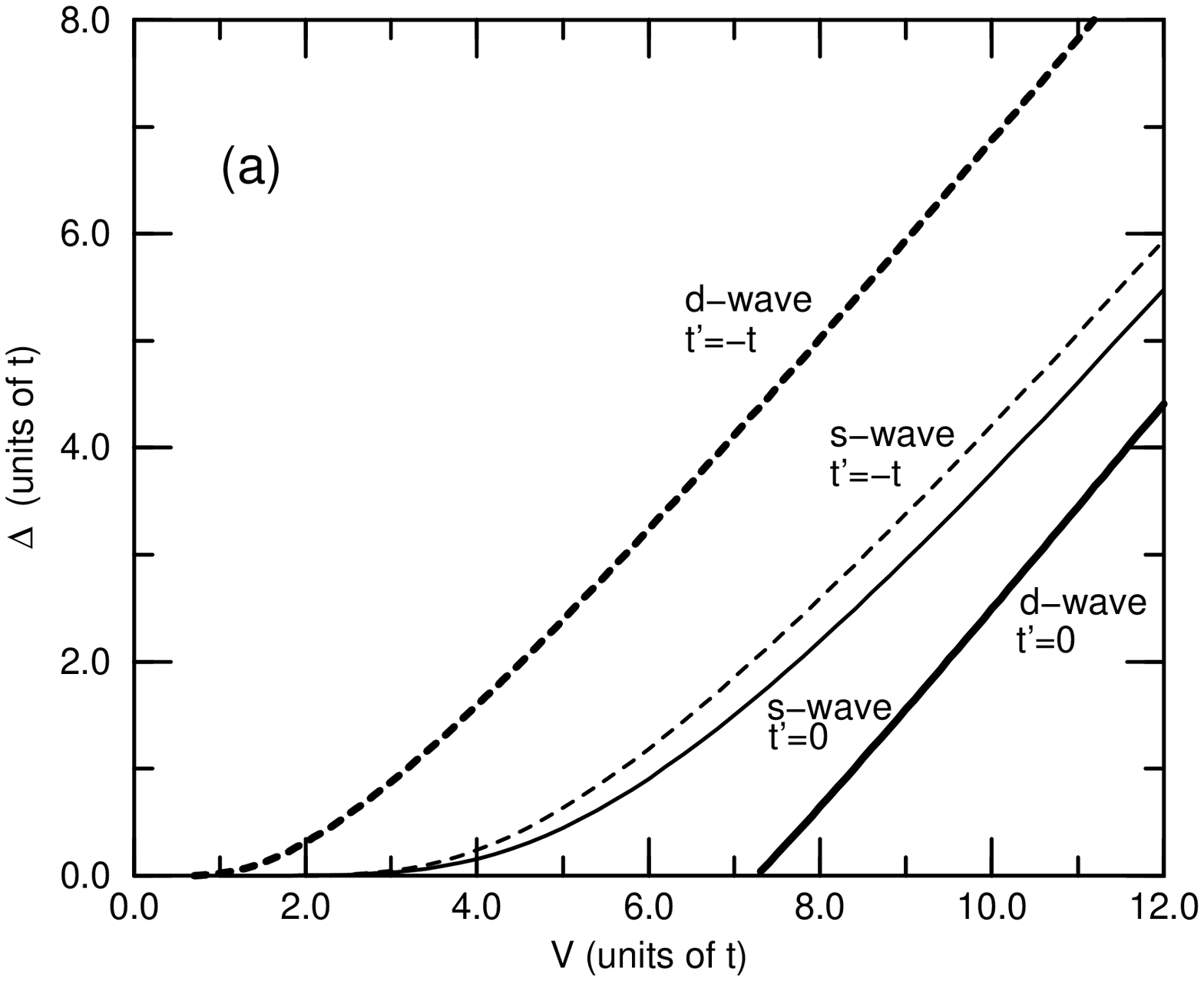} }

\centerline{\epsfxsize=3.0truein \epsfbox{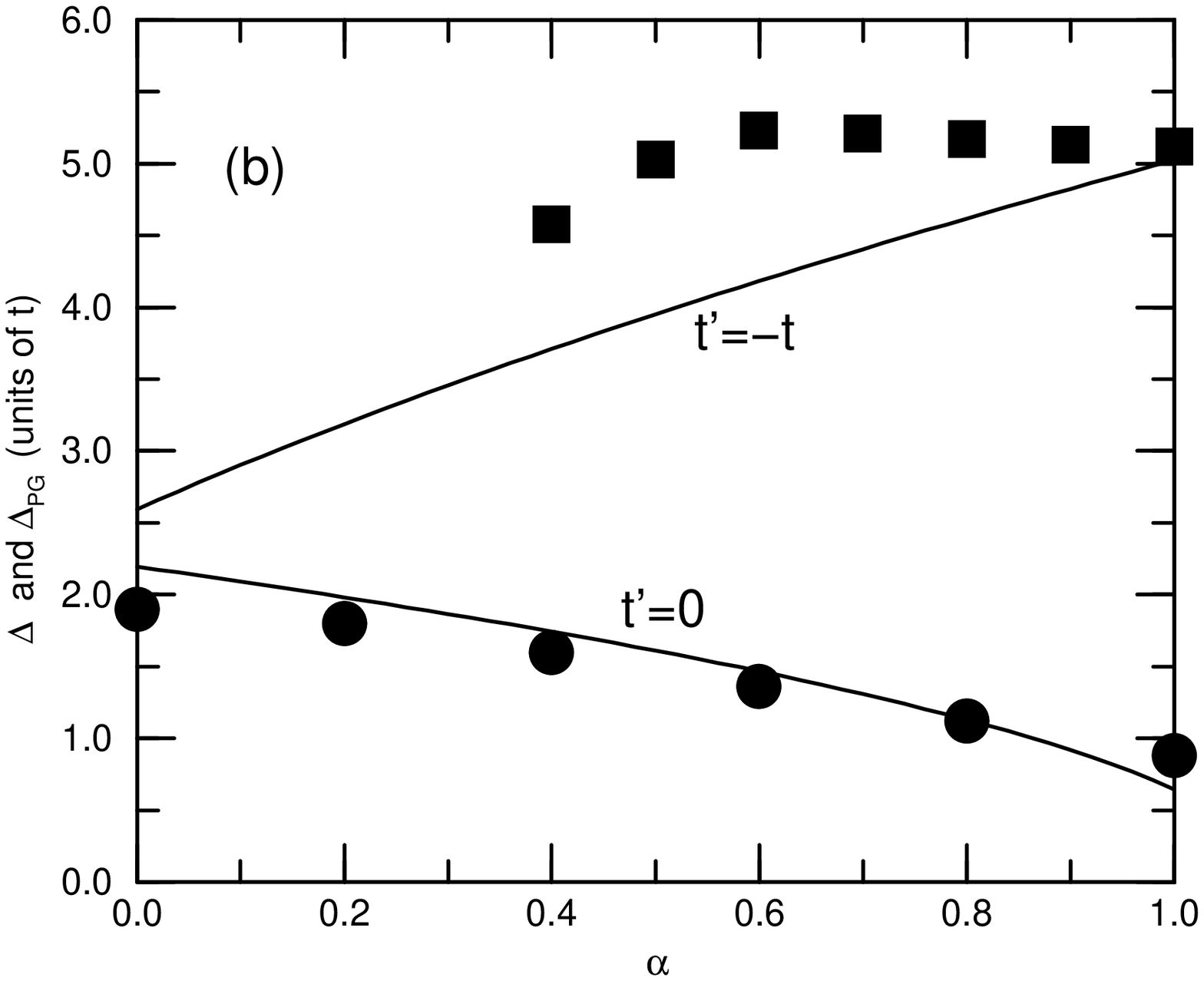} }

\vskip-0.4truein

\caption{
(a) Pair-binding energy as a function of the interaction $V$, which is
an analytic result obtained from the two-particle problem.
(b) Pair-binding energy as a function of $\alpha$ for $t'=0$ (upper solid
curve) and $t'=-t$ (lower solid curve) with $V=8t$. 
The solid symbols indicate the PG
energy $\Delta_{\rm PG}$ estimated from the width between the QP
and the resonant peak in the spectral function.
The solid circle and square indicate $\Delta_{\rm PG}$ for $t'=0$
and $t'=-t$, respectively.
The results at $\alpha=1$ is obtained by the extrapolation from the
results for $\alpha <1$.
}
\label{fig4}
\end{figure}

\section{Comments and summary}
\label{sec:5}

In this paper, pseudogap features in a model for $d$-wave superconductivity 
have been observed.
An important observation to start the discussion is 
that implicitly it has been assumed in the results reported thus far 
that $\Delta_{\rm PG}$ is larger than the superconducting transition 
temperature $T_c$. Otherwise, the results found in our work may be 
confused with the superconducting gap expected below $T_c$.
It is necessary to check this assumption, but it is a very hard task to
calculate the true value of $T_c$.
Then, in order to provide an upper limit for $T_c$, 
the critical temperature is simply evaluated within the mean-field 
approximation.
It is expected that the true $T_c$ will be lower than the mean-field value 
$T_c^{\rm MF}$, which is obtained from the well-known gap equation
\begin{eqnarray}
 	1=V \sum_{\bf k} f_{\bf k}^2 
	{{\rm tanh}[(\varepsilon_{\bf k}-\mu)/(2T_c^{\rm MF})]
	\over \varepsilon_{\bf k}-\mu}.
\end{eqnarray}
In Fig.~5, $T_c^{\rm MF}$ for $d$-wave pairing with
$t'=0$ and $t'=-t$ is shown as a function of $\langle n \rangle$.
For $t'=0$, the calculation for the spectral function  shown in Sec.III
has been done at $\langle n \rangle=0.02$ and $T=0.5t$,
and the point $(\langle n \rangle, T)=(0.02,0.5)$ 
is located above the curve of $T_c^{\rm MF}$ in agreement with our assumption.
Also for $t'=-t$, it is found from the figure
 that the temperature $T=2t$ used for $t' \neq 0$ is larger
than  $T_c^{\rm MF}$, even at $\langle n \rangle=0.2$.
Clearly the temperatures analyzed in the present paper are above the
superconducting critical temperature.
Also note that  $\Delta_{\rm PG}$ for $d$-wave
pairing is larger than $T_c^{\rm MF}$. 
In particular, for the case of $t'=-t$, it is about three times larger 
than $T_c^{\rm MF}$.
This fact clearly suggests the appearance of a pseudogap temperature region, 
$T_c \alt T \alt \Delta_{\rm PG}$, for $d$-wave superconductors models.

\vskip1cm
\begin{figure}[h]
\centerline{\epsfxsize=3.0truein \epsfbox{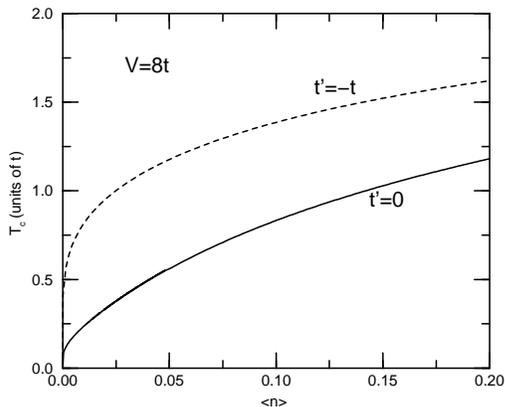} }

\vskip-0.4truein
\caption{Superconducting transition temperature 
in the mean-field approximation for $d$-wave pairing 
as a function of the electron 
number density for $t'=0$ and $t'=-t$ with $V=8t$. 
}
\label{fig5}
\end{figure}

Let us now briefly comment on the imaginary-axis calculation 
with the Pad\'e approximation.
Some attempts were made to obtain the PG structure in the 
imaginary-axis formalism directly for the $d$-wave symmetry,
but it was difficult to observe it in our results.
It might be possible to obtain it, if much more effort was made on the 
imaginary-axis calculations, particularly on the Pad\'e approximation.
However, when the $s$-$d$ conversion trick is also applied to the 
imaginary-axis calculation, a clear sign of the PG 
just below the Fermi level can be easily observed.
Although both results in the real- and imaginary-axis calculations do not 
agree perfectly with each other,
the position of the peak in the imaginary-axis result is 
found to be located just at the lower edge of the PG structure obtained in 
the real-axis calculation.
In the absence of the real-axis results, 
such a small signal of the PG structure may be missed,
because it could be regarded as a spurious result 
due to the Pad\'e approximation. 

Finally, let us discuss the possible relation of our PG
to that observed in the ARPES experiments.
In our result, the PG is characterized by the binding energy of the 
Cooper-pair, which is of the order of $t$ in our models except for a 
numerical factor.
If $t$ is taken as a typical value for HTSC, it becomes of the order of
a sizable fraction of eV, 
which is larger than the observed value in the ARPES experiments.
However, from the viewpoint of the $t$-$J$ model, which is expected to 
contain at least part of the essential physics for the underdoped HTSC, 
the effective hopping is renormalized to be of order $J$, not $t$, 
where $J$($\sim 1000$K) is the antiferromagnetic exchange interaction
between nearest-neighbor spins.\cite{dagotto2}
With this consideration the order of magnitude of our PG energy becomes 
more reasonable.

In summary, the pseudogap structure has been investigated in the low-density 
region for the separable potential model with $s$- as well as $d$-wave 
symmetry. After special technical attention was given to particular features 
of the $d$-wave potential that make some of the calculations unstable,
it has been revealed that the effect on the PG structure of the 
Cooper-pair symmetry manifests in the change of the weight for the 
resonant peak at $A({\bf k^*},\omega)$.
Moreover, it has been clearly shown that the energy scale for the PG 
structure is just the pair binding energy, 
which is certainly larger than $T_c$.

\acknowledgments

The authors thank Alexander Nazarenko for many useful discussions.
T.H. has been supported from the Ministry of Education, Science, 
Sports, and Culture of Japan. E.D. is supported by grant
NSF-DMR-9814350.


\end{multicols}
\end{document}